\documentclass[10pt, 3p,fleqn]{elsarticle}

\usepackage[utf8]{inputenc}
\usepackage{graphics}
\usepackage{graphicx}
\usepackage{amsmath,amssymb,esint}
\usepackage{lineno}
\usepackage{multirow}
\usepackage{subfig}
\usepackage{natbib}
\usepackage{eucal}
\usepackage{mathtools}
\usepackage{placeins}
\usepackage{enumerate}
\usepackage[usenames,dvipsnames]{xcolor}
\usepackage[colorlinks]{hyperref}
\usepackage{setspace}
\usepackage{makecell}
\usepackage{booktabs}
\usepackage{listings}
\usepackage{siunitx}
\usepackage{placeins}
\usepackage{tikz}
\usepackage{pgfplots}
\usepackage{pgfplotstable}
\usetikzlibrary{shapes.geometric, arrows, intersections, through}
\usetikzlibrary{arrows.meta}
\usetikzlibrary{calc}
\tikzset{math3d/.style={z={(-0.65cm,-0.30cm)},y={(0cm,1cm)},x={(0.9cm,-0.15cm)}}}
\usetikzlibrary{shapes.geometric, arrows, intersections, through}
\usetikzlibrary{decorations.text, decorations.shapes, backgrounds}
\usetikzlibrary{arrows.meta}
\usetikzlibrary{colorbrewer}
\usetikzlibrary{patterns}
\usepgfplotslibrary{colorbrewer}
\usepgfplotslibrary{colormaps}

\usepackage{color}
\usepackage[dvipsnames]{xcolor}
\usepackage{soul}

\newcommand{\revtwo}[1]{{\color{black}{#1}}}

\biboptions{sort&compress}
\journal{Journal of Computational Physics}

\pgfplotsset{compat=1.18}

\begin{document}

\begin{frontmatter}

\title{Efficient reduction of vertex clustering using front tracking with surface normal propagation restriction}

\author[1]{Christian Gorges\texorpdfstring{\corref{cor1}}}
\author[2]{Azur Hod\v{z}i\'c}
\author[1,3]{Fabien Evrard}
\author[1]{Berend van Wachem}
\author[2]{Clara M. Velte}
\author[4]{Fabian Denner}

\address[1]{Chair of Mechanical Process Engineering, Otto-von-Guericke-Universit\"{a}t Magdeburg, Universit\"atsplatz 2, 39106 Magdeburg, Germany}
\address[2]{Department of Civil and Mechanical Engineering, Technical University of Denmark, Anker Engelunds Vej 1, Kgs. Lyngby, 2800, Denmark}
\address[3]{Sibley School of Mechanical and Aerospace Engineering, Cornell University, Ithaca, NY 14853, United States of America}
\address[4]{Department of Mechanical Engineering, Polytechnique Montr\'eal, Montr\'eal, H3T 1J4, QC, Canada}

\cortext[cor1]{christian.gorges@ovgu.de}

\begin{abstract}
A significant computational expense and source of numerical errors in front tracking is the remeshing of the triangulated front, required due to distortion and compaction of the front following the Lagrangian advection of its vertices. Additionally, in classic front tracking, the remeshing of the front mesh is required not only due to the deformation of the front shape, but also because the vertices of the front are translated in the direction tangential to the front, induced by the front advection. We present the \textit{normal-only advection} (NOA) front-tracking method with the aim of preventing the tangential motion of the front vertices and the associated vertex clustering, in order to reduce the number of remeshing operations required to retain a high-quality triangulated interface. To this end, we reformulate the velocity used to advect the front at each discrete front-vertex position. The proposed method is validated and tested against the classic front-tracking method, comparing volume conservation, shape preservation, computational costs, and the overall need for front remeshing, as well as experimental results for canonical interfacial flows. The presented results demonstrate that the NOA front-tracking method leads to a typical reduction of remeshing operations by 80 \% or more compared to the classic front-tracking method for well-resolved cases, and results in a smoother front mesh, which is essential for an accurate representation of the geometrical properties of the front. The volume conservation error is reduced by approximately one order of magnitude with the proposed method compared to the classic front-tracking method, at a similar computational cost.
\end{abstract}

\begin{keyword}
    Front tracking \sep
    Interfacial flows \sep
    Remeshing \sep 
    Surface normal velocity\\~\\
    \textcopyright~2023. This manuscript version is made available under the CC-BY-NC-ND 4.0 license. \href{http://creativecommons.org/licenses/by-nc-nd/4.0/}{http://creativecommons.org/licenses/by-nc-nd/4.0/}
\end{keyword}

\end{frontmatter}

\section{Introduction}
\label{sec:Introduction}
The modelling of interfacial flows is a notoriously difficult task. Two aspects of the numerical representation of the interface between two immiscible fluids are particularly challenging: 1)~the accurate transport of the interface through space, and 2)~the appropriate modelling and computation of the force resulting from surface tension. Front tracking \citep{Unverdi1992, Tryggvason2001}, in which a fluid interface in three dimensions (3D) is represented by a triangulated front mesh of which the vertices are advected in a Lagrangian fashion by the underlying velocity flow field, is one of various methods that has been applied successfully to a wide range of interfacial flow problems, see e.g.~\citep{Tryggvason2001, Hua2008, Terashima2010, Muradoglu2014, Irfan2017, Shin2018, Shahin2020}. The major advantage arising from the explicit representation of the interface is the straightforward computation of the normal vector at each point on the interface and, in particular, the estimation of the interface curvature, required for the calculation of the surface tension force. However, as a result of errors in the velocity interpolation and in the time integration, and due to the piecewise-planar nature of the triangulated interface, the advection itself is also not formally conservative~\cite{Tryggvason2001}. The volume conservation and shape preservation errors induced by the velocity interpolation are partly due to the interpolated velocity not being guaranteed to preserve the divergence of the fluid mesh velocity field. Recent works aimed at reducing the velocity interpolation error by applying interpolation methods specifically designed to address this problem \cite{Peskin1993, McDermott2008, Gorges2022}. Furthermore, given that only the discrete surface is advected in front tracking, the local fluid properties, such as density and viscosity, must be re-evaluated at every time instance, and the handling of topology changes is not trivial \cite{Razizadeh2018, Shin2018}.

A significant computational expense and source of numerical errors in front tracking is the remeshing of the triangulated front, which is required due to the distortion and compaction of the front following the Lagrangian advection of its vertices. Additionally, in classic front tracking, remeshing of the front occurs not only due to the deformation of its shape, but also because of the front vertex movement along the surface tangential direction, resulting from the front advection. A well-known example is the simulation of rising bubbles with classic front tracking, illustrated schematically in Figure~\ref{fig:VertexMovement} (see also the video \textit{RisingBubbleClassic.mp4} in the supplementary material). As a response to the surrounding velocity field, the vertices are continually transported from the upstream region of the bubble to its downstream region. In order to retain an adequate resolution and quality of the front mesh, the front-facing area of the front mesh must be continuously refined, whereas the back-facing area must be continuously coarsened.
Another well-known example is the \textit{tank treading} behaviour for front-tracking simulations of droplets or bubbles in shear flow (see the video \textit{ShearFlowClassic.mp4} in the supplementary material), where the front vertices are circulating as a result of the flow on the surface, degrading the mesh quality which in turn requires extensive remeshing. 
While such a Lagrangian motion of the front vertices with the flow simplifies the numerical modeling of the transport of surfactants and other surface-bound quantities, it does not provide any added benefit for the modeling of clean interfaces.
Conventional remeshing operations are error-prone and, strictly speaking, conserve neither the volumes of the interacting bulk phases nor do they preserve the shape of the interface. Mitigating or eliminating these inaccuracies requires additional, and potentially computationally expensive, methods \cite{Gorges2022}. Coarsening algorithms, in particular, tend to affect the volume conservation and shape preservation of the front. The Memoryless-Simplification algorithm of \citet{Lindstrom1998} aims at preserving the local volume during edge collapsing. In our recent work, we proposed a parabolic fit vertex positioning method for edge-splitting and edge-collapsing remeshing operations, which locally approximates the front with a smooth polynomial surface, improving the volume conservation and shape preservation by an order of magnitude compared to conventional remeshing algorithms \cite{Gorges2022}. Furthermore, \citet{deSousa2004} proposed the volume conserving {TSUR3D} algorithm for smoothing small undulations appearing where the front contracts.

\begin{figure}
    \centering
    \includegraphics[scale=0.5]{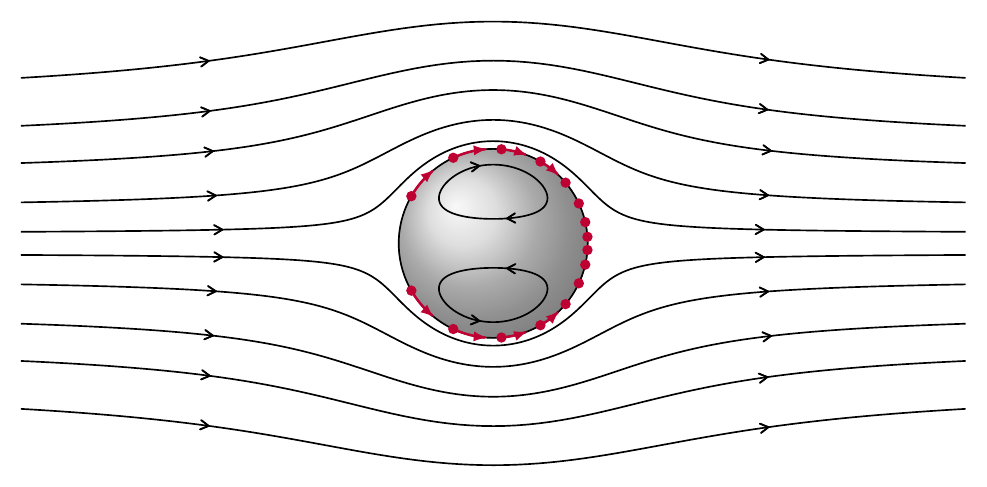}
    \caption{Schematic illustration of the movement of the front vertices (red dots) on a spherical interface subject to a uniform flow, using the classic front-tracking method. The vertices, advected in a Lagrangian fashion, accumulate towards the location of the downstream stagnation point.}
    \label{fig:VertexMovement}
\end{figure}

The objective of the present work is to address the advection and the remeshing of the front using the proposed \textit{normal-only advection} (NOA) front-tracking method, which aims to prevent tangential front-vertex movement and vertex clustering, in order to reduce the number of remeshing operations required to retain a high-quality triangulated interface. To this end, the suppression of vertex movement along the tangential direction of the front is achieved by a reformulation of the surface velocity vector at each discrete front-vertex position. The front vertices are no longer moved with the interpolated fluid velocity as in the classic front-tracking advection, but the reformulated velocity is composed of the sum of the center-of-mass velocity of the body enclosed by the front mesh and a vertex velocity \textit{relative} to the center-of-mass velocity. This relative velocity is decomposed into a surface-tangential and a surface-normal component, from which only the surface-normal velocity component is used for front-vertex advection, thereby cancelling the tangential vertex movement.

The general idea of restricting surface transport to the surface-normal component of the velocity goes back to the beginning of the 20th century with the seminal work of \citet{Hadamard1903}. Recent work from \citet{Grinfeld2012} on the calculus of moving surfaces has led to new applications of the concept to interfacial flow applications. The surface-normal velocity is already used in the field of Arbitrary Lagrangian-Eulerian (ALE) methods for curved and deforming interfaces \cite{Sahu2020, Hirt1974}. Recently, \citet{Deng2022} proposed the mesh-free Moving Eulerian-Lagrangian Particle (MELP) method for simulating incompressible fluids on thin films and foams. The MELP method uses two sets of particles for representing and advecting the interface: a sparse set of Eulerian particles for dynamic interface tracking, and a set of Lagrangian particles for material and momentum transport. Both particle sets move with physical velocities but the Eulerian components are exclusively advected with surface-normal velocities~\cite{Deng2022}. 
In the context of front tracking, the idea of advecting the front using only the the surface-normal velocity has been used in relation to modelling evaporation and mass transfer in two-dimensional flows \citep{Juric1998, Tryggvason2015, Irfan2017}. To this end, the velocity due to mass transfer, which is by definition oriented normal to the surface, in combined with the local fluid velocity, which may also be projected onto the surface normal direction \citep{Irfan2017}. The NOA front-tracking method proposed in the current work extends the use of the surface-normal velocity to the advection of the front in three-dimensional flows and, by introducing a suitable reference velocity, generalizes the use of the surface-normal velocity to achieve a significant reduction in remeshing operations. The NOA method is tested on canonical cases, such as rising bubbles in a quiescent flow and a droplet in shear flow, focusing on volume conservation, shape preservation and the computational costs.

This article is structured as follows: In section \ref{section:Numerical framework}, we describe the numerical framework in which the novel NOA front-tracking method is embedded. In section \ref{section:NOA}, the normal-only advection is then derived in detail, followed by section \ref{section:Results}, in which the proposed method is validated and compared to the classic front-tracking method with respect to the total number of remeshing operations, volume conservation, shape preservation and computational costs. In section \ref{sec:Limitations}, we discuss the limitations of the NOA front-tracking method and propose a hybrid approach for reducing numerical errors at high curvature regions. Section \ref{section:Summary and conclusions} provides a summary and conclusions.

\section{Numerical framework}
\label{section:Numerical framework}

Using the numerical framework of the front-tracking method to simulate incompressible interfacial flows, a divergence-free flow field governed by the Navier-Stokes equations is solved
\begin{align}
    \nabla \cdot \mathbf{u} &= 0 \, ,\\
    \rho \left( \frac{\partial \mathbf{u}}{\partial t} + \nabla \cdot ( \mathbf{u} \otimes \mathbf{u}) \right) &= - \nabla p + \nabla \cdot \boldsymbol{\tau} + \rho \mathbf{g} + \mathbf{f}_{\sigma} \delta_\text{S} \,,
\end{align}
using the one-fluid formulation on the Eulerian mesh, with $\rho$ the fluid density, $\mathbf{u}$ the velocity vector, $p$ the pressure, $\boldsymbol{\tau}$ the viscous stress tensor, $\mathbf{g}$ the gravity acceleration vector, $\mathbf{f}_{\sigma}$ the surface tension force, and $\delta_\text{S} = \delta_\text{S} (\mathbf{x} - \mathbf{x}_\text{S})$ the interfacial delta function. The incompressible Navier-Stokes equations are discretised and solved using a finite-volume framework with collocated variable arrangement, which solves the equations in a coupled, pressure-based manner with second-order accuracy in space and time \cite{Denner2014a, Denner2020}.

The physical fluid properties, such as density and viscosity, are approximated using a smooth indicator function, $\mathcal{I}(\mathbf{x})$. This indicator function is constructed based on the discrete front mesh position and assumes the value $1$ in the Eulerian mesh cells which only contain phase '$\text{A}$' and $0$ in the Eulerian mesh cells which only contain the phase '$\text{B}$'. A general fluid property, $\phi$, is calculated as $\phi (\mathbf{x}) = \phi_{\text{B}} + (\phi_{\text{A}} - \phi_{\text{B}}) \, \mathcal{I}(\mathbf{x})$.
The indicator function is computed by spreading the indicator gradient from the front mesh onto the fluid mesh by means of the Peskin interpolation kernel \citep{Peskin2003,Peskin1972} and, subsequently, solving a Poisson equation, as detailed in \citep{Tryggvason2001, Tryggvason2011a}.

The surface tension force is defined as $\mathbf{f}_{\sigma} = \sigma \kappa \mathbf{n}$,
where $\sigma$ is the surface tension coefficient, $\kappa$ the curvature and $\mathbf{n}$ the normal vector of the surface. The surface tension force on a triangulated front element $E$ can be computed by using the Frenet-Element method given as  \cite{Tryggvason2001, Tryggvason2011a} 
\begin{equation}
    \mathbf{F}_{\sigma, E} = \sigma \int_{E} \kappa \mathbf{n} \mathrm{d}A = \sigma \oint_C \mathbf{p} \mathrm{d}l = \sigma \sum\limits_{e} \mathbf{p}_e l_e\,, 
\end{equation}
where, with the help of the Stokes theorem, the area integral of element $E$ with its boundary $C$ is converted into a line integral along the edges $e$ of the front triangle.
The subscript $e$ refers to the edges of the triangle with length $l_e$ and planar vector $\mathbf{p}_e = \mathbf{n}_e \times \mathbf{t}_e$ as the cross product of the edge normal vector $\mathbf{n}_e$ and edge tangential vector $\mathbf{t}_e$. Accordingly, the surface tension force acting on the triangles of the front is distributed onto the fluid mesh as
\begin{equation}
    \mathbf{f}_{\sigma}(\mathbf{x}) = \sum_E \mathbf{F}_{\sigma, E} D(\mathbf{x} - \mathbf{x}_E)\,,   
\end{equation}
where the distribution function $D(\mathbf{x} - \mathbf{x}_E)$ is given by the Peskin interpolation kernel \citep{Peskin2003,Peskin1972}.

The front-tracking method and the governing equations are not only coupled via the indicator function and the surface tension force, but also by the interpolation of the fluid velocity to the positions of the front vertices for their advection. The Peskin interpolation kernel is also used to interpolate the velocity in the course of this study, but different interpolation methods may be used in conjunction with the normal only advection method proposed in Section \ref{section:NOA}.

\section{Normal-only advection (NOA)}
\label{section:NOA}

In front tracking, the vertices of the discrete front are advected in a Lagrangian fashion as $\mathrm{d}\mathbf{x}_{i} (t)/\mathrm{d}t = \mathbf{u}(\mathbf{x}_{i} (t), t)$,
where $\mathbf{x}_{i}$ is the position vector of the $i$-th vertex of the front and $\mathbf{u}$ is the fluid velocity at the vertex position $\mathbf{x}_{i}$. Because the fluid velocity is only known at discrete points on the fluid mesh, the velocity at the vertex position must be interpolated from the discrete points of the fluid mesh. In practice, this leads to an advection equation at each point in time of the form
\begin{equation}
    \frac{\mathrm{d}\mathbf{x}_{i}}{\mathrm{d}t} = \mathbf{\bar{u}}(\mathbf{x}_{i} (t), t)\,,
    \label{VertexAdvection2}
\end{equation}
where $\mathbf{\bar{u}}$ indicates an interpolated fluid velocity.

The surface velocity $\mathbf{u}$ can be rewritten in terms of a sum of a reference velocity $\mathbf{u}_{\text{ref}}$, for instance a volume-averaged velocity of the body enclosed by the front, and a vertex velocity relative to the reference velocity $\mathbf{u}_{\text{rel}}$, such that $ \mathbf{u} = \mathbf{u}_{\text{ref}} + \mathbf{u}_{\text{rel}}$,
with
\begin{equation}
    \mathbf{u}_{\text{ref}} = \frac{1}{V_{\text{d}}} \int_{\Omega} \mathbf{u} \ \mathrm{d}\mathbf{x}\,,
    \label{Eq:V_m}
\end{equation}
and
\begin{equation}
    \mathbf{u}_{\text{rel}} = \mathbf{u} - \mathbf{u}_{\text{ref}}\,.
    \label{Eq:V_r}
\end{equation}
Here, $V_{\text{d}}$ is the volume of the body $\Omega$ enclosed by the surface and $\mathbf{u}$ is the corresponding fluid velocity. 
By applying this methodology to the discrete front-tracking case, equations \eqref{Eq:V_m} and \eqref{Eq:V_r} become
\begin{equation}
  \mathbf{u}_{\text{ref}} \approx \frac{\sum_P \mathbf{u}_P \mathcal{I}_P V_P}{\sum_P\mathcal{I}_P V_P}\,, \label{eq:uref_volume}
\end{equation}
and $\mathbf{u}_{\text{rel}} \approx \bar{\mathbf{u}} - \mathbf{u}_{\text{ref}}$,
respectively, where $P$ denotes all cells of the fluid mesh under the assumption of having only one interface-bounded body in the domain. In front tracking, the tangential component of the relative velocity $\mathbf{u}_{\text{rel}}$ is typically causing the vertices to be inhomogeneously spread on the interface. Therefore, the aim of the proposed NOA method is to project the relative velocity  $\mathbf{u}_{\text{rel}}$ onto the surface normal direction, so that the tangential motion of the vertices is suppressed, which, in turn, reduces vertex clustering and postpones or in some cases eliminates entirely the necessity of front remeshing. 

\begin{figure}
    \centering
    \includegraphics[scale=0.6]{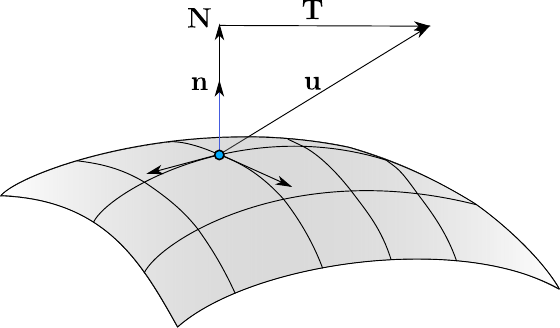}
    \caption{Velocity vector decomposition on a surface into a normal and tangential component. The normal vector of the surface is denoted with $\mathbf{n}$.}
    \label{fig:Velocity_decomposition}
\end{figure}

As shown in Figure \ref{fig:Velocity_decomposition}, the surface velocity can be decomposed into a normal component and a tangential component as
\begin{equation}
    \mathbf{u} = \underbrace{(\mathbf{u} \cdot \mathbf{n}) \, \mathbf{n}}_{\text{normal}} + \underbrace{\phantom{(}\mathbf{P} \cdot \mathbf{u}\phantom{)}}_{\text{tangential}},
\end{equation}
where $\mathbf{P} = \mathbf{I} - \mathbf{n} \otimes \mathbf{n}$ is the surface projection tensor and $\mathbf{I}$ is the identity tensor. 
The normal component of the relative velocity follows as
\begin{equation}
    \mathbf{u}_{\text{rel,n}} = (\mathbf{u}_{\text{rel}} \cdot \mathbf{n})\,\mathbf{n}  \label{Eq:Vrel_n}
\end{equation}
and, together with the reference velocity, see equation \eqref{eq:uref_volume},
the \textit{normal-only advection} (NOA) velocity for the front-vertex advection is given as
\begin{equation}
    \frac{\mathrm{d}\mathbf{x}_{i} (t)}{\mathrm{d}t} = \mathbf{u}_{\text{NOA}} (\mathbf{x}_{i}(t),t) = \mathbf{u}_{\text{ref}} (t) + \mathbf{u}_{\text{rel,n}} (t).
    \label{Eq:NOA-Advection}
\end{equation}
The advection of the front vertices using a surface-normal velocity proposed in previous studies \citep{Juric1998, Irfan2017} for the simulation of interfacial flows with evaporation and mass transfer is a special case of NOA, with $\mathbf{u}_\mathrm{ref} = 0$.

For the time integration of equation \eqref{Eq:NOA-Advection}, the fourth-order Runge-Kutta scheme is used in this study.
The vertex normal vector and $\mathbf{u}_{\text{rel,n}}$ are updated at each step of the Runge-Kutta scheme, whereas $\mathbf{u}_{\text{ref}}$ is computed only at $t$ and assumed to be constant for the steps of the Runge-Kutta scheme in each time step. 

\subsection{Computation of the vertex normal vector}
The surface normal vector is required at the vertices of the front for the decomposition of the velocity vector described in equation \eqref{Eq:Vrel_n}. In this study, a weighted sum of the adjacent triangle normals is considered for the calculation of the vertex normal vector, which is computed as $\mathbf{n}_{v} = \sum_{t} w_{t} \mathbf{n}_t$,
$w_{t} = \alpha_{t} / L_{t}$.
The subscripts ${v}$ and ${t}$ denote a vertex and a triangle of the front, respectively, $\alpha_{t}$ is the angle between the two edges of the current triangle that share the vertex under consideration and $L_{t}$ is the product of the lengths of the two edges of the current triangle that share the vertex under consideration. 

\subsection{Addressing surface undulations}
In regions where the front vertices tend to accumulate, undulations on the front mesh may occur due to variations in the velocity field from one fluid cell to the next. This non-physical behaviour introduces errors in the computation of the geometric front properties, and is accompanied by a degradation of an accurate front advection and surface tension computation. Therefore, to ensure that the front mesh is free of spurious undulations for computing its geometric properties, the undulation-removal algorithm {TSUR3D} proposed by \citet{deSousa2004} is applied. 

\subsection{Volume correction}
\label{sec:Volume correction}
A major issue in front tracking is volume conservation. The most common reasons for a lack of volume conservation are the use of interpolation methods that do not preserve the divergence of the underlying velocity field, low-order time integration schemes used for the advection and inaccurate remeshing operations. Regardless of the cause, the volume conservation error at each time step is small but cumulative, which can cause significant volume errors for long simulations. \citet{Pivello2012} introduced a volume-correction step after the front advection, assuming that the volume changes within each time-step are small and only occur because of expansion or shrinkage in the direction normal to the front. Since the proposed NOA front-tracking method advects the front vertices exclusively in the direction normal to the front, this volume correction step is expected to harmonize well with the proposed method and is, therefore, applied in the further course of this study.

By assuming that the volume change is constant over the front mesh, the volume change associated with a single front triangle is given by $\Delta V_t = A_t \, h$ \cite{Pivello2012},
with $h$ the length of the mean expansion (or shrinkage) in the normal direction and $A_t$ the triangle area. Hence, the total volume change for the body enclosed by the front is computed as $\Delta V_\Omega = A_{\Omega} \, h$,
with $A_{\Omega}$ denoting the surface area of the front mesh, given as the sum of all triangle areas. By rearranging this equation, 
the length $h$ in the normal direction between the initial and current volume is $h = \Delta V_\Omega/A_{\Omega}$,
and the volume correction can be achieved by shifting the positions of the front vertices by factor $h$ in the normal direction, $ \mathbf{x}_{\text{corr}} = \mathbf{x}_{\text{init}} - h \, \mathbf{n}_{v}$.
The subscripts '$\text{corr}$' and '$\text{init}$' denote the volume-corrected vertex position and the initial vertex position right after the advection of the front, respectively.

\section{Validation and results}
\label{section:Results}
The proposed NOA front-tracking method is validated, tested, and compared to the classic front-tracking method in relation to two categories of test cases: 1) rising bubbles in a quiescent fluid, and 2) droplets in shear flow.

The parameters for the rising bubble are given in Table \ref{tab:RisingBubbleSetups}. The expected terminal Reynolds number $\mathrm{Re} = \rho_c u_{\text{T}} d_{\text{b}}/\mu_\text{c}$,
with $u_\text{T}$ the terminal rise velocity and $d_\text{b}$ the initial bubble diameter, is increasing from case R1 with  $\mathrm{Re} = 0.232$, \textit{i.e.}~Stokes flow, to R4 with $\mathrm{Re} = 55.3$. The Bond number
$\mathrm{Bo} = \rho_\text{c} g d_\text{b}^2/ \sigma$,
and the Morton number $\mathrm{Mo} = g \mu_\text{c}^4 / (\rho_\text{c} \sigma^3)$,
are varied throughout the four test cases and are characterizing the terminal shape of the rising bubbles. The density and viscosity ratios, the initial ratio of the bubble diameter to fluid mesh spacing, $d_\text{b} / \Delta x$, and the ratio of the domain size to initial bubble diameter, $D / d_\text{b}$, are the same for all four cases. Guided by the work of \citet{Hua2008}, the domain size and fluid mesh spacing are chosen such that they do not have an appreciable influence on the results. A no-slip boundary condition is imposed at the bottom of the domain, a pressure outlet is defined at the top of the domain, and periodic boundary conditions are imposed at the side walls. The dimensionless characteristic time for the rising bubble is defined as $ \tau = t \, \sqrt{g/d_\text{b}}$.

\begin{table}[t]
    \centering
    \caption{Setup of the four considered rising bubbles adopted from \citet{Bhaga1981}.}
    \label{tab:RisingBubbleSetups}
    \begin{tabular}{ p{3.5cm}||p{2cm}|p{2cm}|p{2cm}|p{2cm}  }
 Settings & R1 & R2 & R3 & R4\\
 \hline
 \hline
 Reynolds number & 0.232 & 7.77 & 18.3 & 55.3\\
 \hline
 Bond number & 17.7 & 243 & 339 & 32.2\\
 \hline
 Morton number & 711 & 266 & 43.1 & \num{8.2e-4}\\
 \hline
 $d_\mathrm{b} / \Delta x$ & \multicolumn{4}{c}{20}\\
 \hline
 $D / d_\mathrm{b}$ & \multicolumn{4}{c}{8}\\
 \hline
 $\rho_\mathrm{c} / \rho_\mathrm{d}$ & \multicolumn{4}{c}{1000}\\
 \hline
 $\mu_\mathrm{c} / \mu_\mathrm{d}$ & \multicolumn{4}{c}{100}
\end{tabular}
\end{table}

\begin{figure}
\begin{center}
  \includegraphics[scale=0.5]{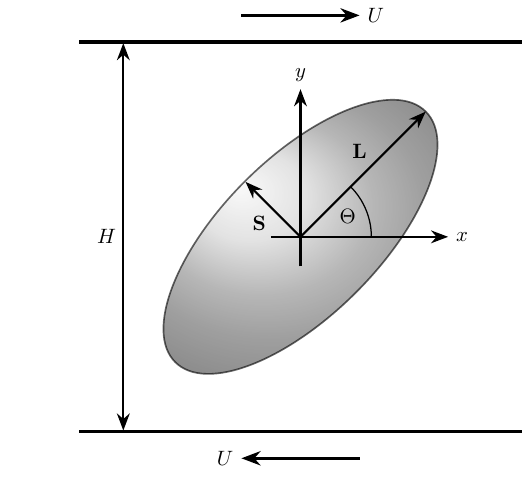}
    \caption{Schematic illustration of the droplet in shear flow. The height of the domain in the considered test case is $H=5 \, r_\mathrm{d}$.}
    \label{fig:sheardrop}
\end{center}
\end{figure}

The second test scenario is a droplet in shear flow, for which some basic geometric properties are depicted in Figure \ref{fig:sheardrop}. Important parameters in this case are the capillary number $\mathrm{Ca} = r_\text{d} \mu_\text{c} \gamma/\sigma$,
with $r_\text{d}$ the initial radius of the droplet and $\gamma =2U/H= 15 \, \text{s}^{-1}$ the applied shear rate of the flow, as well as the viscosity ratio and the density ratio. For the validation with the results from \citet{Feigl2007}, the capillary number is varied from $0.1$ to $0.3$. Since the deformation of the droplet is relatively small for $\mathrm{Ca} \leq 0.3$, the capillary number is increased to $\mathrm{Ca} = 0.6$ for a more meaningful comparison of the proposed NOA front-tracking method with the classic front-tracking method. The viscosity ratio is $0.3$ and the density ratio is unity for this test case. The domain has the dimensions $10 r_\text{d} \times 5 r_\text{d} \times 10 r_\text{d}$. The mesh spacing of the fluid mesh is $\Delta x = r_\text{d} / 9$ and the dimensionless characteristic time for the droplet in shear flow is defined as $\tau = t \sigma/(r_\text{d} \mu_\text{c})$.
The shape of the droplet is characterized by the inclination angle $\Theta$ and the Taylor deformation parameter $D = |\mathbf{L}|-|\mathbf{S}|/(|\mathbf{L}|+|\mathbf{S}|)$,
%
%
where $\mathbf{L}$ and $\mathbf{S}$ are the vectors representing the major and minor axes of the droplet, respectively, as illustrated in Figure \ref{fig:sheardrop}.

Using these test cases, we evaluate the proposed NOA front-tracking method in terms of (i) the accuracy of the advection of the front, (ii) the reduction of remeshing operations applied to the front, (iii) the accuracy and the representation of geometric properties of the interface using the front mesh, and (iv) the computational costs.
For all considered test cases, the velocity interpolation from the fluid mesh to the front vertices by the Peskin interpolation, the computation of the indicator function and the surface tension force, the volume-correction step, the method to remesh the front and the application of the {TSUR3D} algorithm are the same for both the proposed NOA front-tracking method and the classic front-tracking method. The front is remeshed by edge-splitting and edge-collapsing, whereby the vertices are positioned using a parabolic fit, as well as edge-flipping, as described in detail in \cite{Gorges2022}. At each time step, the remeshing algorithm checks if the front mesh consists of edges which need to be remeshed and the remeshing algorithm is applied if necessary. The {TSUR3D} algorithm is applied in a frequency such that undulations of the front mesh are suppressed to a sufficient degree, while ensuring that the qualitative front deformation is minimized, every 10th to 20th time step in the presented simulations. The time step applied to simulate the rising bubbles and the droplet in shear flows satisfies the capillary time step constraint \cite{Denner2015}, $\Delta t \leq \sqrt{(\rho_\text{A} + \rho_\text{B}) \Delta x^3/ (2 \pi \sigma)}$.

\subsection{Validation}
The NOA front-tracking method is validated against the experimental results of \citet{Bhaga1981} for rising bubbles in different flow regimes (see Table \ref{tab:RisingBubbleSetups}) as well as against the results of \citet{Feigl2007} for droplets in shear flow. Furthermore, the results of the proposed NOA front-tracking method are compared to the results of the classic front-tracking method. 

Table \ref{tab:RisingBubbleReynoldsnumbers} provides the terminal Reynolds numbers of the four rising bubble cases considered for the NOA front-tracking method, the classic front-tracking method, as well as the experimental results of \citet{Bhaga1981}. The terminal Reynolds numbers obtained with NOA front tracking and classic front tracking are in very good agreement. Additionally, the terminal Reynolds numbers of the simulations are in good agreement with the experiments from \citet{Bhaga1981} as well as numerical results for these bubbles previously reported with other front-tracking implementations \cite{Hua2008,Pivello2014}.

\begin{figure}
    \centering
    \includegraphics[scale=0.7]{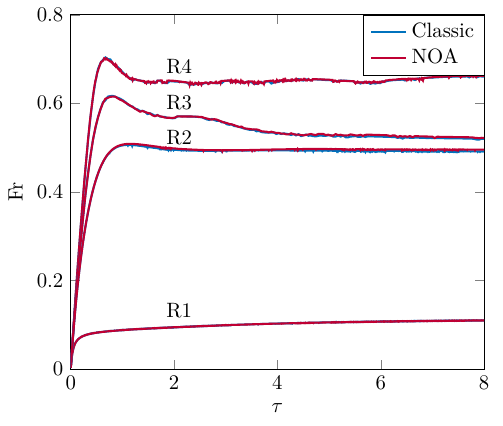}
    \caption{Evolution of the Froude number $\mathrm{Fr}$ as a function of the dimensionless time $\tau$ for the four rising bubbles, obtained with the proposed NOA front-tracking method and the classic front-tracking method.}
    \label{fig:FroudeAll}
\end{figure}

\begin{figure}
    \centering
    \subfloat[R1, NOA front tracking]{\includegraphics[width=0.249\textwidth]{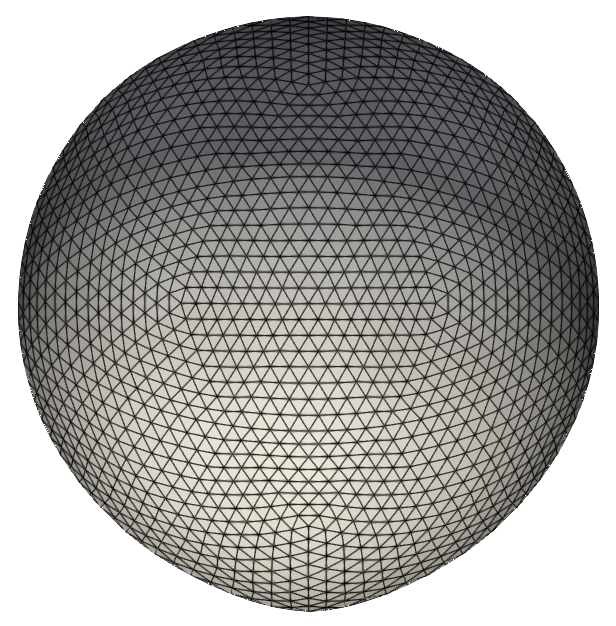}} \hfill
     \subfloat[R2, classic front tracking]{\includegraphics[width=0.249\textwidth]{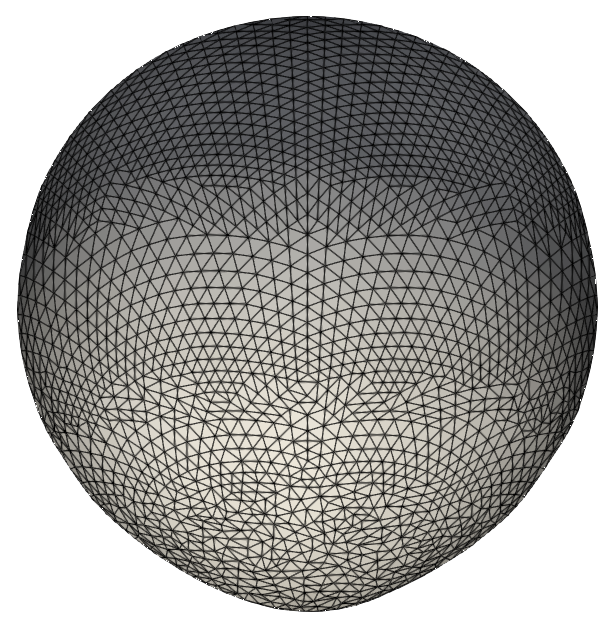}} \hfill
\subfloat[R2, NOA front tracking]{\includegraphics[width=0.249\textwidth]{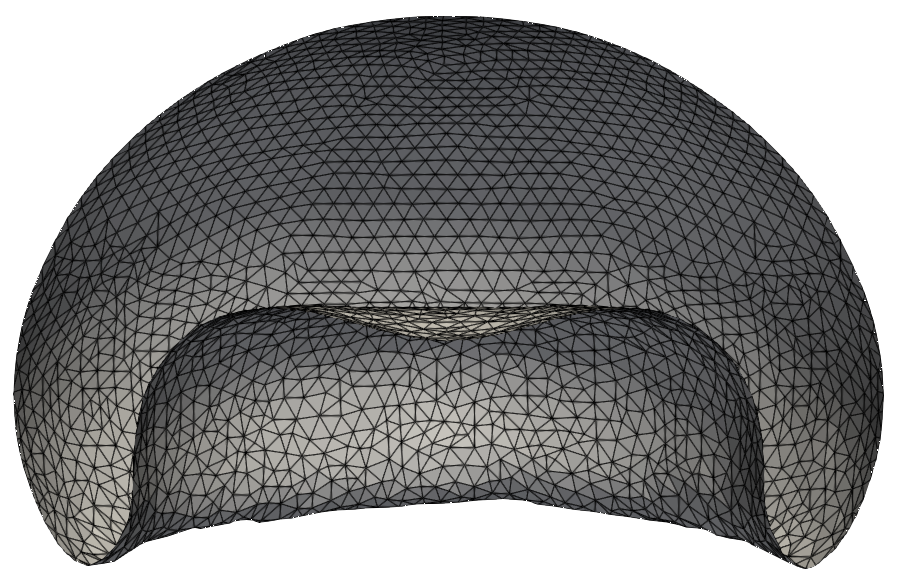}} \hfill
     \subfloat[R2, classic front tracking]{\includegraphics[width=0.249\textwidth]{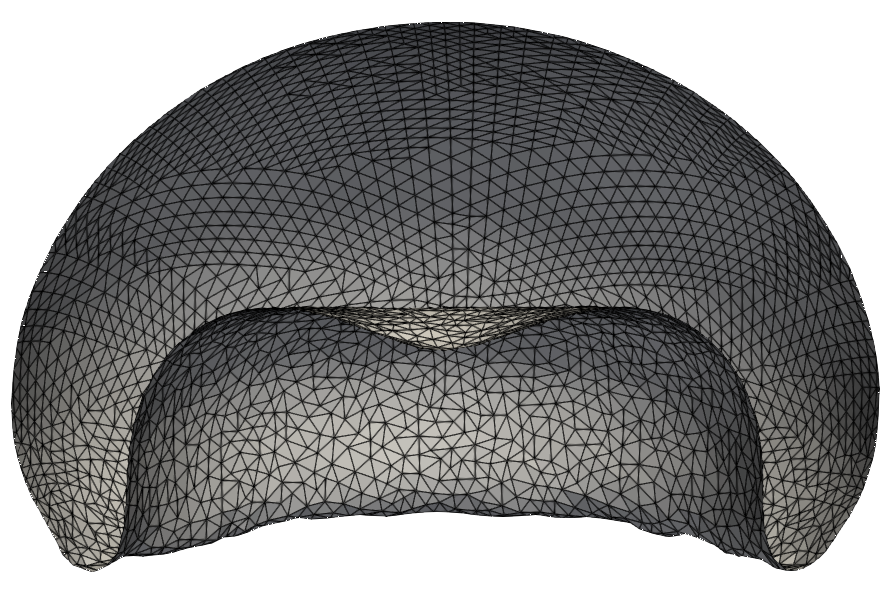}}\\
     \subfloat[R3, NOA front tracking]{\includegraphics[width=0.249\textwidth]{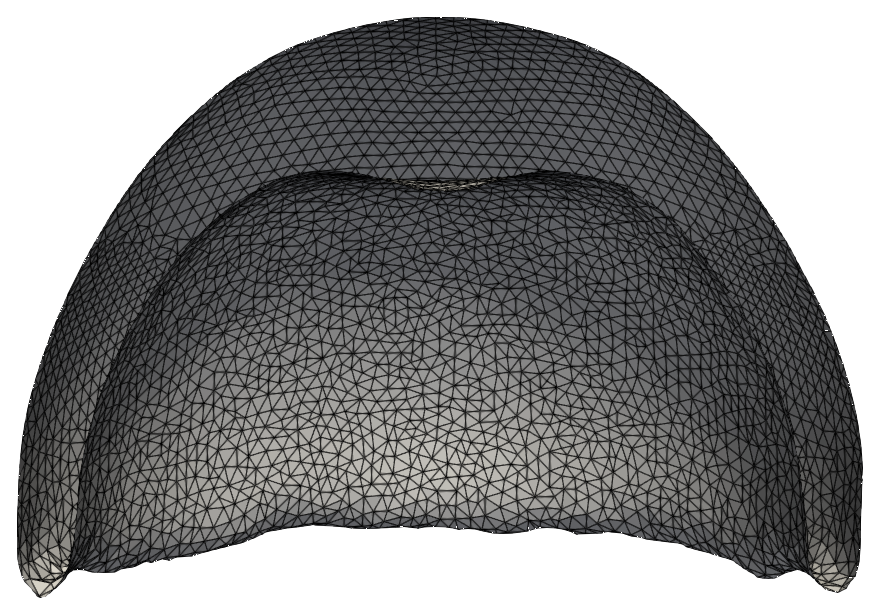}} \hfill
     \subfloat[R3, classic front tracking]{\includegraphics[width=0.249\textwidth]{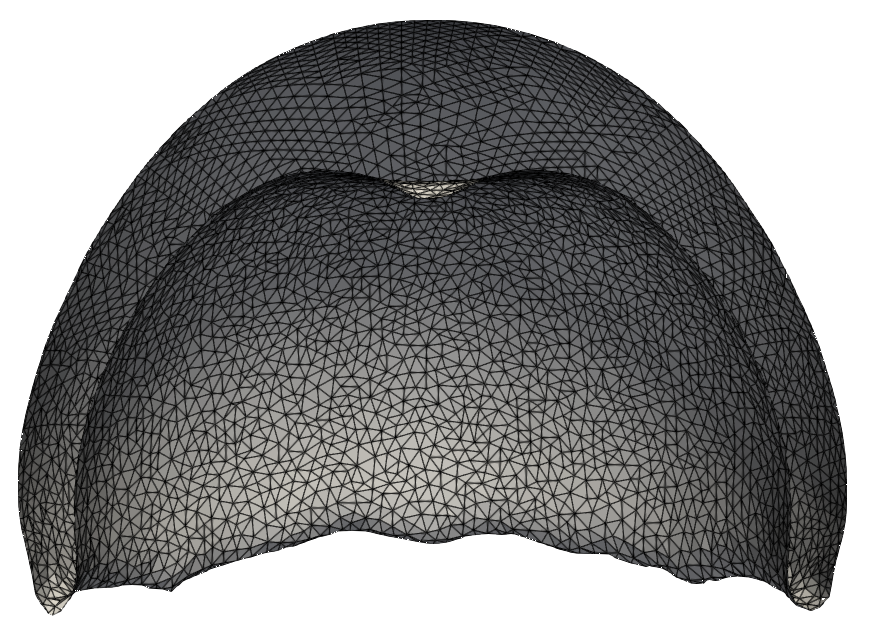}}\hfill
\subfloat[R4, NOA front tracking]{\includegraphics[width=0.249\textwidth]{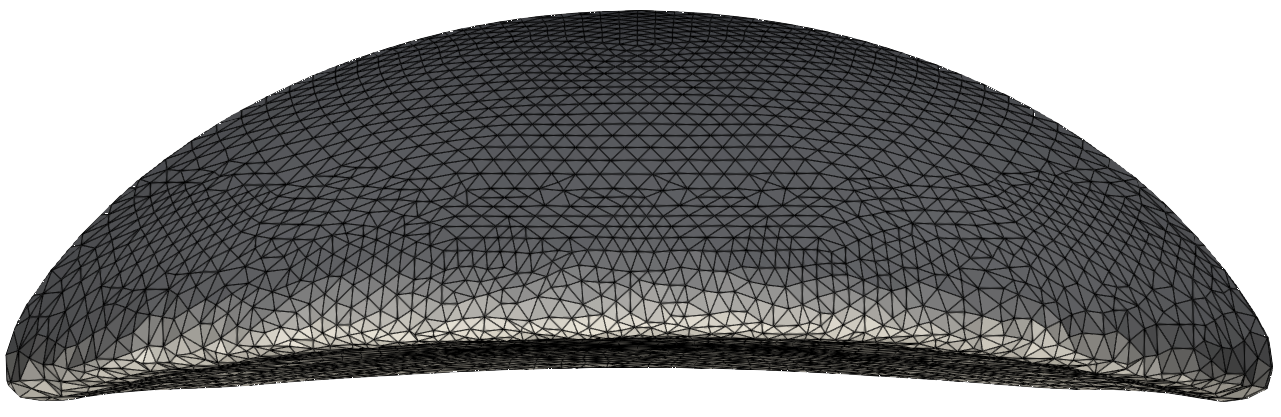}}\hfill
     \subfloat[0.49\textwidth][R4, classic front tracking]{\includegraphics[width=0.249\textwidth]{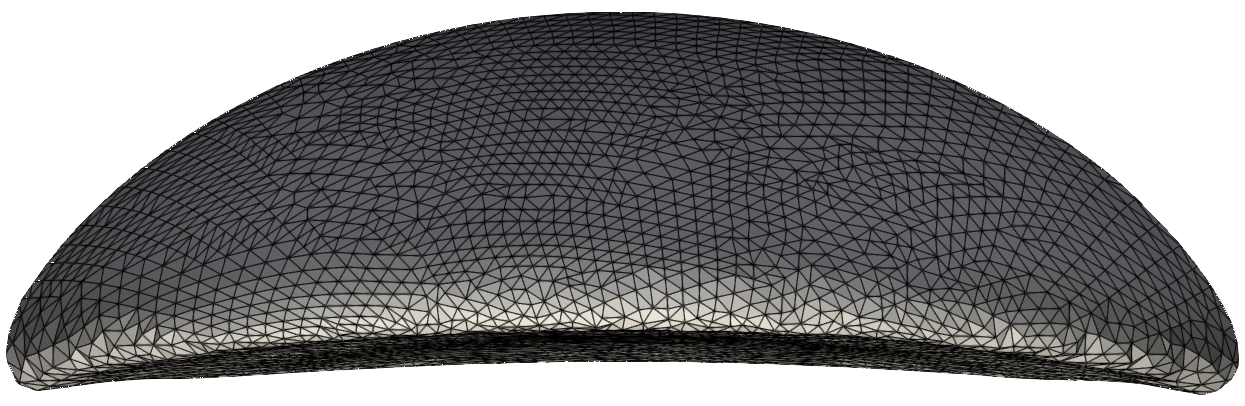}}\\
     \subfloat[Shear flow, NOA front tracking]{\includegraphics[width=0.279\textwidth]{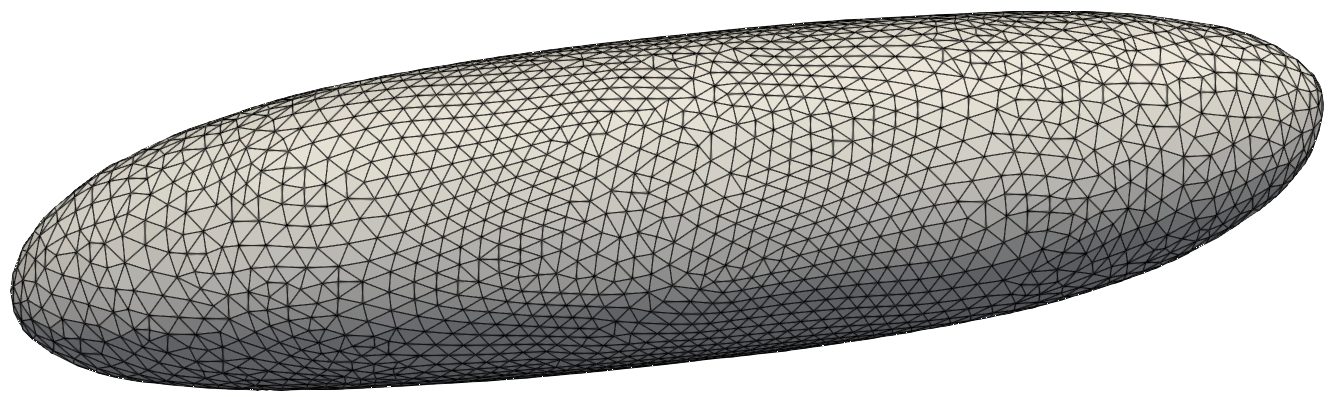}} \quad 
     \subfloat[Shear flow, classic front tracking]{\includegraphics[width=0.279\textwidth]{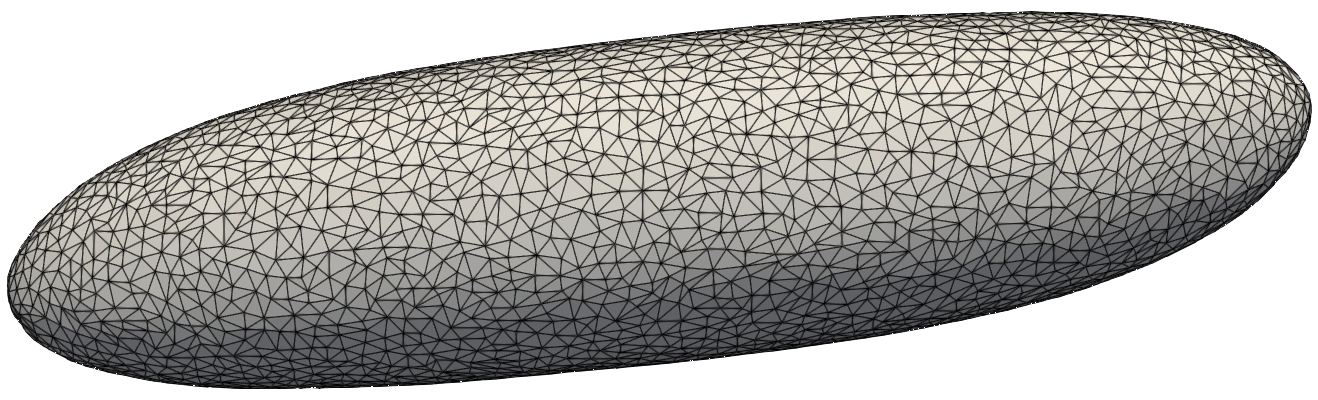}}
    \caption{Comparison of the front meshes obtained with the NOA front-tracking method (left) and the classic front-tracking method (right) for the terminal shapes of the rising bubbles and the droplet in shear flow.}
    \label{fig:TerminalShapes}
\end{figure}

Figure \ref{fig:FroudeAll} shows the evolution of the Froude number $\mathrm{Fr} = \sqrt{u_\mathrm{T}/(g \, d_\mathrm{b})}$,
%
%
as a function of dimensionless time for the four considered rising bubbles. The Froude numbers obtained with both, the proposed NOA front-tracking method and the classic front-tracking method, are in very close agreement throughout the simulations, demonstrating that NOA advects the interface with an accuracy comparable to classic front tracking. Figure \ref{fig:TerminalShapes} also compares the terminal shapes of the rising bubbles for both front-tracking methods, which are in good qualitative agreement with each other and, for instance, with numerical results previously reported in the literature \cite{Hua2008, Pivello2014}.

\begin{figure}
    \centering
    \includegraphics[scale=0.7]{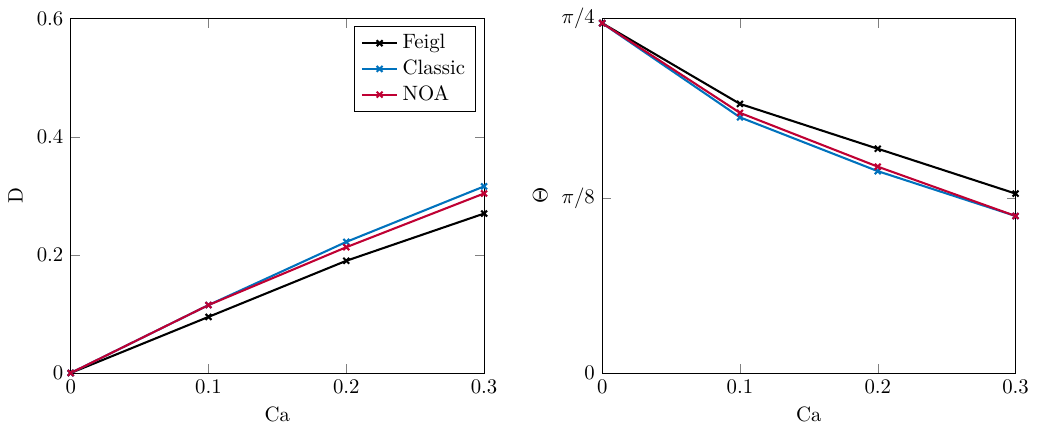}
    \caption{Taylor deformation parameter $D$ and inclination angle $\Theta$ for a droplet in shear flow using various capillary numbers $\mathrm{Ca}$ for the validation of the NOA and classic front-tracking methods with simulations from \citet{Feigl2007} for $\mu_\mathrm{c} / \mu_\mathrm{d} = 0.3$.}
    \label{fig:ShearFlowValidation}
\end{figure}

The validating results for the droplets in shear flow are shown in Figure \ref{fig:ShearFlowValidation}. Both front-tracking methods are in very good agreement with each other and in reasonably good agreement with the simulations of \citet{Feigl2007}. Additionally, Figure \ref{fig:TerminalShapes} illustrates the deformation of the droplet predicted by both front-tracking methods for a capillary number of $\mathrm{Ca} = 0.6$. The NOA front-tracking method predicts a virtually identical shape of the droplet as the classic front-tracking method.

\begin{table}
    \centering
    \caption{Terminal Reynolds numbers of the rising bubbles obtained using the NOA front-tracking method, the classic front-tracking method and measured in the experiments of \citet{Bhaga1981}.}
    \label{tab:RisingBubbleReynoldsnumbers}
    \begin{tabular}{ p{1.3cm}||p{1.7cm}|p{1.7cm}|p{2.5cm}  }
      & \multicolumn{3}{c}{Re}\\
     \hline
     Case & NOA FT & Classic FT & Bhaga \& Weber\\
     \hline
     \hline
      R1 & 0.195 & 0.195 & 0.232\\
      R2 & 7.63 & 7.59 & 7.77\\
      R3 & 16.9 & 16.8 & 18.3\\
      R4 & 54.4 & 53.9 & 55.3\\
    \end{tabular}
\end{table}


\begin{table}[ht]
    \centering
    \caption{Comparison of the number of remeshing operations (edge-splitting, edge-collapsing, edge-flipping) per time step for the NOA front-tracking method and the classic front-tracking method in the unsteady and steady flow regimes.}
    \label{tab:RemeshingOperations}
    \begin{tabular}{ p{2cm}p{2cm}||p{1.7cm}|p{1.7cm}|l}
      & & \multicolumn{3}{c}{Remeshing operations per time step}\\
     \hline
     Case & & NOA FT & Classic FT & Change with NOA FT\\
     \hline
     \hline
     \multirow{ 2}{*}{R1} & unsteady & 0.00 & 2.34 & \\
      & steady & 0.00 & 3.34 & \\
      \hline
      \multirow{ 2}{*}{R2} & unsteady & 3.93 & 43.26 & -91 \%\\
      & steady & 3.93 & 51.74 & -92 \%\\
      \hline
      \multirow{ 2}{*}{R3} & unsteady & 12.51 & 64.10 & -80 \%\\
      & steady & 9.76 & 69.03 & -86 \%\\
      \hline
      \multirow{ 2}{*}{R4} & unsteady & 15.48 & 104.44 & -85 \%\\
      & steady & 6.84 & 112.52 & -94 \%\\
      \hline
      \multirow{ 2}{*}{Shear flow} & unsteady & 0.24 & 1.34 & -82 \%\\
      & steady & 0.07 & 2.00 & -97 \%\\
     
    \end{tabular}    
\end{table}

\subsection{Remeshing operations}
\label{sec:remeshingops}
\revtwo{The main motivation for the development of the NOA front-tracking method is to reduce the number of required remeshing operations. To highlight the differences between the NOA and classical front-tracking methods, the results for the number of remeshing operations are divided into the unsteady flow regime and the steady flow regime. The unsteady flow regime corresponds to the part of the simulation in which the interface shape changes, while the steady part corresponds to the part of the simulation in which the interface no longer changes significantly. The unsteady regime for all test cases is defined as the interval from $\tau = 0$ to $\tau = 7.5$. In this regime the total number of remeshing operations (edge-splitting, edge-collapsing, edge-flipping) is incremented until $\tau = 7.5$ and divided by the number of time steps to obtain an averaged number of remeshing operations per time step. The steady regime starts at $\tau = 7.5$. To facilitate a direct comparison between both regimes, the remeshing operations in the steady regime are counted for the same number of time steps as in the unsteady regime. Table \ref{tab:RemeshingOperations} shows the average number of remeshing operations per time step for the steady and unsteady regimes of the corresponding test cases with NOA and classical front-tracking.}

For test case R1, NOA front tracking is able to eliminate the need for remeshing entirely, since the bubble hardly deforms as it rises with a terminal Reynolds number below unity. Classic front tracking, however, still requires more than \revtwo{2 remeshing operations per time step} to retain an adequate resolution and quality of the front mesh for case R1. This difference in remeshing throughout the evolution of the front mesh is highlighted in Figure \ref{fig:TerminalShapes} and the videos \textit{RisingBubbleNOA.mp4} and \textit{RisingBubbleClassic.mp4} in the supplementary material. In the classic front-tracking method the mesh has to be continuously refined at the top and coarsened at the bottom, because of the tangential motion of the front vertices from the top of the bubble to its bottom. In contrast, the front vertices using NOA do not move in the tangential direction of the interface and, consequently, the front does not need to be remeshed. 
Taking the reference velocity to be $\mathbf{u}_{\text{ref}} = 0$, instead of the volume-averaged velocity defined in equation \eqref{eq:uref_volume}, as considered in previous studies \citep{Juric1998, Irfan2017}, does not benefit from the reduction in remeshing operations and for test case R1 \revtwo{requires on average more than 4 remeshing operations per time step}. A suitably chosen reference velocity is, thus, key to reducing the number of remeshing operations. A comparison of all three variants can also be seen in the supplementary material video \textit{RisingBubbleComparison.mp4}.

Since the shape of the remaining three bubbles considered is changing significantly as they rise, the front must also be remeshed when using NOA front tracking, in order to retain an adequate mesh resolution and quality. However, significantly fewer remeshing operations are required using NOA than for the classic front tracking. Figure \ref{fig:TerminalShapes} illustrates that using NOA, remeshing is only applied in areas where the front expands or contracts, with very few remeshing operations applied at the top of the bubbles. As a result, the triangle density at the top of the bubbles obtained with the proposed method is more uniform compared to the triangle density at the top of the bubbles obtained with the classic front-tracking method.

Regarding the droplet in shear flow, NOA prevents tank treading, whereby the front vertices move with the flow in a circular manner along the tangential direction of the droplet surface. The front mesh is only remeshed at the tips and the center of the droplet, where the droplet is stretched or contracted as a result of the shear flow. The classic front-tracking method, however, requires constant remeshing to account for the tangential vertex movement as a result of the tank treading motion of the droplet. This is illustrated in the videos \textit{ShearFlowNOA.mp4} and \textit{ShearFlowClassic.mp4} in the supplementary material. The improvement in the mesh quality, in terms of uniformity and the shape of the triangles, achieved by NOA is clearly visible in Figure \ref{fig:TerminalShapes}.

\revtwo{Table \ref{tab:RemeshingOperations} shows that the average number of remeshing operations for the classical front-tracking is overall slightly higher in the steady regime compared to the unsteady regime. This is counterintuitive, since the interface shape deformation is larger in the unsteady regime and, therefore, the number of remeshing operations should be higher in the unsteady regime. This phenomenon can be explained by two reasons. First, the deformation of the interface shape does not start immediately at the first time step, but it takes a short time for the interface motion to start, and second, the numerical deformation caused by the tangential motion of the Lagrangian vertices does not cease in the steady regime and, therefore, depending on the strength of the deformation in the unsteady regime, the number of remeshing operations in the steady regime remains approximately the same. The comparison of the number of remeshing operations for the NOA front-tracking method in the two regimes shows that the number of remeshing operations decreases significantly in the steady regime compared to the unsteady regime. Comparing the NOA front-tracking method with the classical front-tracking method, we find that the number of remeshing operations is reduced by at least 80 \% in both regimes and for all test cases.}

\subsection{Volume conservation}
\begin{figure}
    \centering
    \includegraphics[scale=0.7]{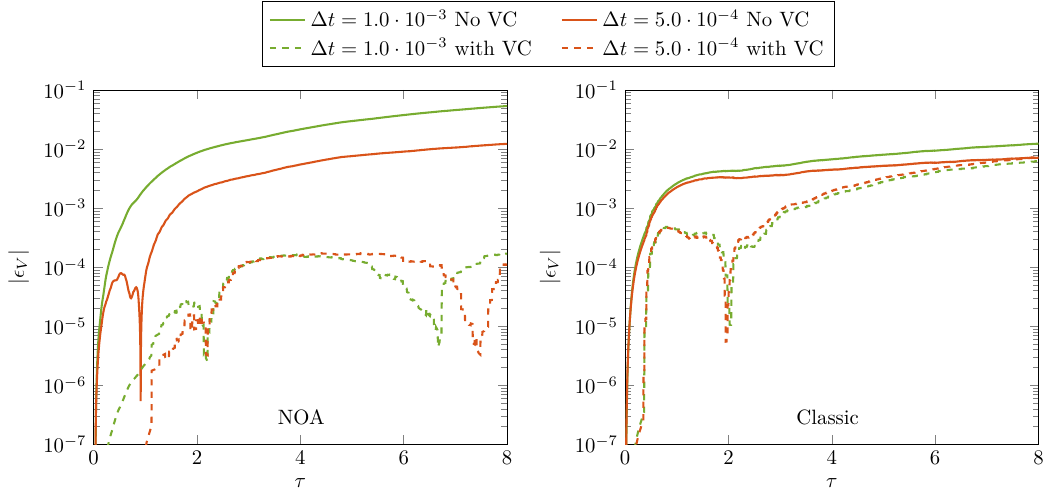}
    \caption{Evolution of the volume error over dimensionless time and the corresponding impact of the time step size $\Delta t$ and the volume correction step (VC) for the proposed NOA front-tracking method (left) and the classic front-tracking method (right) for the R2 test case.}
    \label{fig:VolumeErrorTimeStep}
\end{figure}

Figure \ref{fig:VolumeErrorTimeStep} shows the evolution of the volume error for the rising bubble R2 using NOA and the classic front-tracking method. The figure shows results for two different time steps, with and without the volume-correction step discussed in Section \ref{sec:Volume correction}. The volume correction reduces the volume error by approximately two orders of magnitude using the NOA front-tracking method, whereas the applied volume correction only has a minor influence on the volume error using the classic front-tracking method. 
Comparing the volume error between methods without the volume correction shows that the volume error of the classic front-tracking method is lower than the volume error using the NOA front-tracking method. However, in conjunction with the volume-correction step, the volume error using NOA is more than one order of magnitude smaller than the volume error obtained with the classic front-tracking method. 
Furthermore, applying the volume-correction step evidently reduces the dependency of the volume error on the applied time step.

These results indicate that the overall volume error for the NOA front-tracking method is dominated by errors arising directly from the advection of the front, which is explained by the positive influence of the volume-correction step. For the classic front-tracking method, the volume-correction step only has a minor influence on the overall volume error, which may indicate that the volume error in classic front tracking is dominated by the errors arising from the remeshing operations.

\begin{figure}[ht]
    \centering
    \includegraphics[scale=0.7]{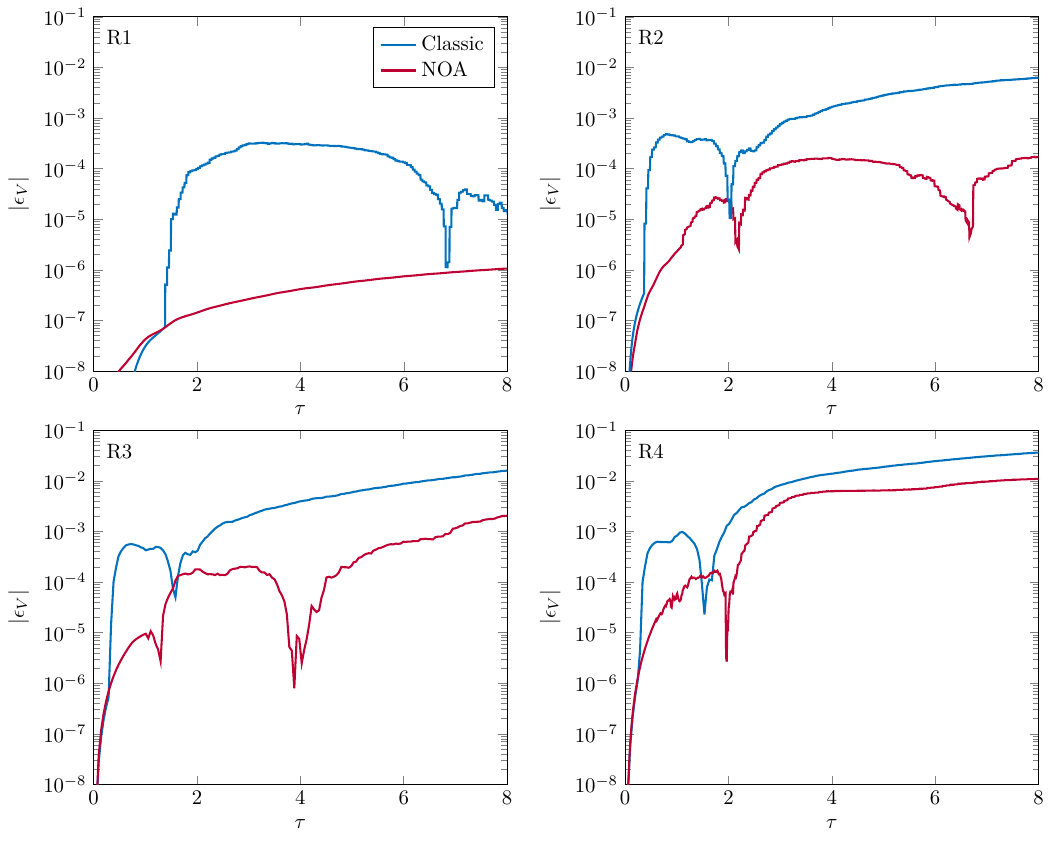}
    \caption{Evolution of the volume error over dimensionless time for the four considered rising bubbles.}
    \label{fig:VolumeErrorAll}
\end{figure}

\begin{figure}[ht]
    \centering
    \includegraphics[scale=0.7]{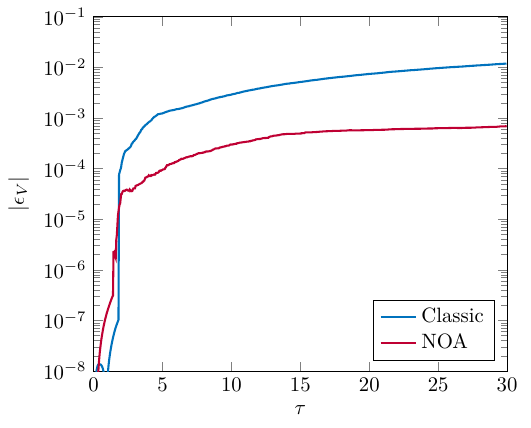}
    \caption{Evolution of the volume error over dimensionless time for the considered droplet in shear flow.}
    \label{fig:VolumeErrorShearFlow}
\end{figure}

Figures \ref{fig:VolumeErrorAll} and \ref{fig:VolumeErrorShearFlow} show the volume error for the four rising bubbles and the droplet in shear flow, respectively. The final volume error using NOA is in all considered cases smaller than with classic front tracking, ranging from a difference of approximately factor 3 for the R4 case to approximately factor 30 for the R2 case. Over time, the volume error using NOA is almost exclusively smaller than the volume error obtained with the classic front-tracking method. For the rising bubble test cases it can be seen that the magnitude of the volume error increases with increasing terminal Reynolds number.


\subsection{Shape preservation}
For an accurate computation of the geometric properties of the interface, the preservation of the shape of the front is essential. As an indicator for the quality of the shape preservation, the mean curvature $\bar{\kappa}$ averaged over all triangles of the front and the corresponding normalized standard deviation $\sigma / \bar{\kappa}$ are chosen for the rising bubble test case with the lowest Reynolds number (case R1) and the droplet in shear flow. 

\begin{figure}
    \centering
    \includegraphics[scale=0.7]{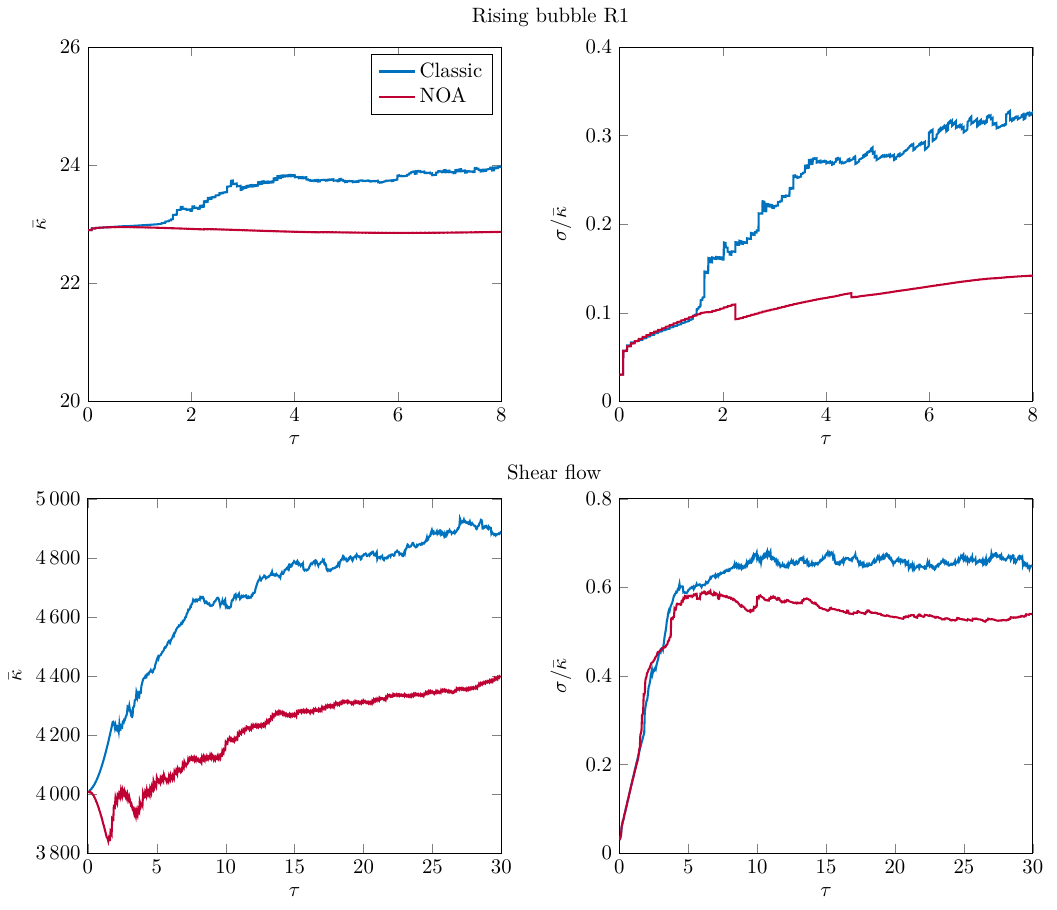}
    \caption{Evolution of the mean curvature averaged over all front triangles, $\bar{\kappa}$, and the corresponding normalized standard deviation, $\sigma / \bar{\kappa}$, over dimensionless time for rising bubble R1 ($\mathrm{Re}=0.232$) and the droplet in shear flow.}
    \label{fig:Curvature_STDW}
\end{figure}

The evolution of the average mean curvature and the normalized standard deviation for case R1 are shown in Figure \ref{fig:Curvature_STDW}. Since the bubble is rising in the Stokes regime, the shape of the bubble remains nearly spherical throughout the simulation. For a perfect sphere, the mean curvature of the surface is constant and its standard deviation is zero. The left graph in Figure \ref{fig:Curvature_STDW} shows that the mean curvature obtained using NOA remains nearly constant, as expected for a rising bubble in the Stokes regime. The right graph shows that the normalized standard deviation of the mean curvature is approximately 0.15 at time $\tau=8$. In contrast, the mean curvature and the corresponding normalized standard deviation obtained with the classic front-tracking method increase continuously, mostly because of remeshing, with the standard deviation approaching approximately 0.33 at $\tau=8$.

The standard deviation of the mean curvature can serve as an indicator of the smoothness of the front mesh. An increasing roughness of the front or undulations on the front mesh, due to remeshing operations, are typically non-physical and are an important factor in an accurate computation of the geometric properties of the front and for conservation of the volume enclosed by the front. Even though the remeshing operations used in this work employ a parabolic fit to position the front vertices more accurately \cite{Gorges2022}, which significantly improves the preservation of the shape of the front, remeshing still has a negative impact. As can be seen in Figure \ref{fig:Curvature_STDW} for case R1, the proposed NOA front-tracking method considerably decreases the negative impact associated with remeshing by simply reducing the number of required remeshing operations.

Figure \ref{fig:Curvature_STDW} also shows the evolution of the average mean curvature and its normalized standard deviation for the droplet in shear flow. Given the large deformations of the droplet, the average mean curvature can merely provide a basis for the computation of the standard deviation. The standard deviation obtained using NOA is significantly smaller than the standard deviation obtained with the classic front-tracking method. This indicates that the front mesh of the proposed method is smoother than the front mesh of the classic front-tracking method, which allows for more accurate and robust computations of the geometric properties.

\subsection{Computational cost}
The difference in computational cost required for the unit operations associated with the advection and remeshing of the front is shown in Table \ref{tab:ComputationalTime} for the considered test cases, \revtwo{which are divided into the unsteady and steady regimes. The same procedure as for the averaging of the remeshing operations per time step is applied to compute the average cost per time step for the advection and remeshing unit operations in the unsteady and steady regimes.} It is noteworthy that the advection of the front is the only unit operation which is explicitly modified by the NOA front-tracking method. Changes in the computational cost spent for remeshing using the NOA method compared to classic front-tracking is an indirect consequence of the judicious advection of the front vertices by the former.

\revtwo{In general, the advection of the front is computationally more expensive using the NOA method, because of the additional computation of the velocity of the center of mass of the enclosed volume and of the vertex normal vectors. The reduction in the computational cost for the R1 test case using NOA is the result of a smaller number of vertices that are advected in the course of the simulation compared to the classic front-tracking method that has to continuously remesh the front as a result of the surface tangential vertex motion.}

Given the overall reduction of the number of remeshing operations by eliminating surface-tangential vertex motion, the computational cost spent on remeshing unit operations is always smaller using NOA than using the classic front-tracking method. \revtwo{However, even though the total number of remeshing operations is reduced by at least 80 \% for all test cases (see Table \ref{tab:RemeshingOperations}), the difference in computational cost also depends on the shape of the front mesh.} For cases in which the front mesh deforms more, NOA requires an increased number of remeshing operations of the front mesh, even though the total number of remeshing operations is smaller than with classic front tracking. The smaller computational cost difference results from the significant number of iterations over the entirety of the front mesh in relation to a single remeshing operation. The number of front edge splittings/collapses is then only of small importance in regard to the computational cost.
With respect to the total computational cost spent on advecting and remeshing the front, listed in the fourth column of Table \ref{tab:ComputationalTime}, the computational costs of the NOA method are smaller than those of the classic front-tracking method, for the considered test cases. 
Test cases with a moderate change in the shape of the front strongly shift the cost advantage to the NOA method.

\begin{table}[t]
    \centering
    \caption{Difference in computational cost per time step spent on advection and remeshing unit operations with the NOA front-tracking method compared to the classic front-tracking method for selected test cases in the unsteady and steady flow regimes.}
    \label{tab:ComputationalTime}
    
    \begin{tabular}{ p{2cm}p{2cm}||p{2cm}|p{2cm}|p{4cm}}
     & & \multicolumn{3}{c}{Computational cost per time step}\\
     \hline
 Case & & Advection & Remeshing & Advection + Remeshing\\ 
 \hline
 \hline
  \multirow{ 2}{*}{Shear flow} & unsteady & $+20$ \% & $-31$ \% & $+1$ \% \\
  & steady & $+17$ \% & $-40$ \% & $-3$ \% \\
  \hline
  \multirow{ 2}{*}{R1} & unsteady & $-8$ \% & $-81$ \% & $-26$ \% \\
  & steady & $-31$ \% & $-87$ \% & $-47$ \% \\
  \hline
  \multirow{ 2}{*}{R2} & unsteady & $+3$ \% & $-2$ \% & $+2$ \% \\
  & steady & $+1$ \% & $-5$ \% & $-3$ \% \\

\end{tabular}
\end{table}

\section{Limitations and hybrid approach}
\label{sec:Limitations}
The NOA front-tracking method has been developed with the aim of reducing the tangential vertex motion in simulations of physically relevant interfacial flows. Nonetheless, one of the most widely used test scenarios for interface advection methods is the artificial three-dimensional interface deformation in a divergence-free velocity field of \citet{LeVeque1996}. This test case is designed to test the limits of interface advection methods by strong deformations, inducing large interface curvatures.

\begin{figure}
    \centering
    \includegraphics[scale=0.68]{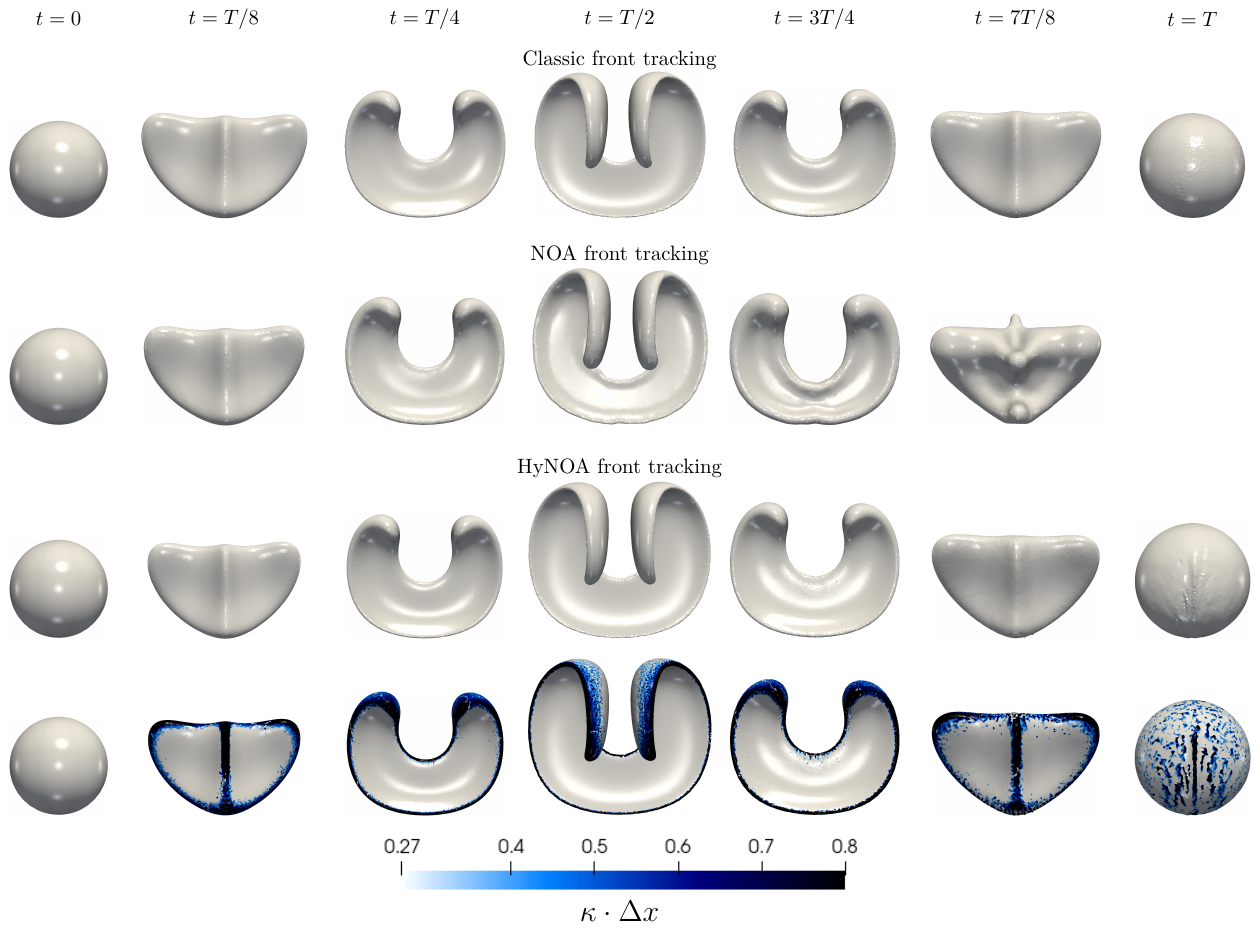}
    \caption{Evolution (from left to right) of the sphere in the deforming velocity field of \citet{LeVeque1996}. Up to $t = T/2$ the initially spherical front is stretched and afterwards (under ideal conditions) returns to its initial position and shape. The color in the bottom figure shows the normalized mean curvature.}
    \label{fig:LeVeque}
\end{figure}

The initially spherical interface is deformed by an analytically defined velocity field, given as
\begin{subequations}
\begin{eqnarray}
u(\mathbf{x}, t) &= 2 \sin^2 \left(\pi x \right) \sin \left( 2\pi y \right) \sin \left( 2\pi z \right) \cos \left( \pi t/T \right)\,,\\
v(\mathbf{x}, t) &= -\sin\left( 2\pi x \right) \sin^2 \left( \pi y \right) \sin \left( 2\pi z \right) \cos \left( \pi t/T \right)\,,\\
w(\mathbf{x}, t) &= -\sin \left( 2\pi x \right) \sin \left( 2\pi y \right) \sin^2 \left( \pi z \right) \cos \left( \pi t/T \right)\,.
\end{eqnarray}
\end{subequations}
The velocity field is time dependent and applied on the time interval $0\leq t \leq T$, where $T$ is set {\it a priori}. Up to $t=T/2$ the front is strongly deformed and for $T/2 < t \leq T$ the velocity field is reversed, such that under ideal conditions, the initial position and shape of the initial sphere are recovered at $t=T$. 
To focus on the functionality and capabilities of the NOA method in this test case, the exact analytical velocity at the front vertices is used as the total velocity, without interpolating the velocity to the front. The errors introduced by the advection and remeshing of the front are, consequently, the only computational operations that can cause differences between the start and end positions of the front. 
The way in which the interface is deformed depends largely on its initial position. Figure \ref{fig:LeVeque} shows the evolution of the deforming front initially centered at $\mathbf{x} = [0.35 \ 0.35 \ 0.35]$ with diameter $d=0.3$, inside a unit cube for $T = 3.0$. The ratio of the  initial diameter to the mean triangle edge length is $d/\bar{l}_{e} \approx 53$, with the initial value of $\bar{l}_e$ also used as the reference length scale for the remeshing of the front \cite{Gorges2022}. The mesh spacing of the fluid mesh for reconstructing the indicator function by means of solving a Poisson equation is $d/\Delta x = 15$.

Figure \ref{fig:LeVeque} illustrates that the extreme deformations of the interface in this test case expose the limitations of the NOA front-tracking method at surface locations characterized by very large mean curvatures. In these locations the surface normal vector estimation is less accurate, leading to a deterioration of the front quality. For $t \geq 7T/8$, the front breaks up due to filament formation. The deteriorating quality of the numerical estimation of the interface normal vector in areas of large mean curvature adversely affects the advection of the front using the NOA method, where the interface normal vector is used to determine the velocity of the front vertices.

In order to improve the robustness of the NOA method, a hybrid approach (HyNOA) combining the NOA method with the classic front-tracking method is proposed. In this case, the NOA method is applied where the normalized mean curvature is $\kappa \, \Delta x < 0.8$, while the classic front-tracking method is applied where $\kappa \, \Delta x \geq 0.8$. The front regions advected with the NOA front-tracking method and those advected with classic front-tracking can be seen from Figure \ref{fig:LeVeque}: the black regions of the interface ($\kappa \, \Delta x > 0.8$) are advected with classic front tracking, whereas the majority of the front is advected with the NOA method. The final shape of the interface obtained with HyNOA is slightly more undulated compared to its initial geometry, and is comparable to the final interface shape using the classical front-tracking method. As the surface normal estimation errors increase, surface undulations are generated using the NOA method. Although the {TSUR3D} algorithm is tailored to overcome such undulations, its frequent use must necessarily be limited in order to avoid a contrived front geometry. The deterioration of the front shape in the high curvature regions using the NOA method, most clearly illustrated at $t = T/2$ and $t = 3T/4$, is, thus, partly caused by the {TSUR3D} algorithm. Applying the {TSUR3D} algorithm with a small frequency, however, adversely affects the accuracy of the estimation of surface normal vectors and, in turn, the advection of the front. 

\section{Conclusions}
\label{section:Summary and conclusions}
The \textit{normal-only advection} (NOA) front-tracking method has been proposed with the aim of reducing the vertex clustering of the triangulated front representing the fluid interface and thereby reducing the number of remeshing operations required to retain a high-quality triangulated interface. Fundamental to the NOA method is the suppression of the surface tangential motion. This is achieved by decomposing the velocities interpolated onto the front into the center-of-mass velocity and vertex velocities relative to the center-of-mass velocity. The surface normal components of the latter are then used to restrict the movements of the surface vertices in the direction normal to the front. 

The proposed method has been validated and tested against the classic front-tracking method as well as compared to experimental results for canonical interfacial flows, such as rising bubbles and a droplet in shear flow. Furthermore, the proposed method was compared to the classic front-tracking method in terms of volume conservation, shape preservation, computational costs, and the overall need of front remeshing. Results show that the NOA front-tracking method leads to a reduction of remeshing operations by up to 100 \% compared to the classic front-tracking method, and results in a smoother front mesh, which is essential for an accurate representation of the geometrical properties of the front. The volume conservation error is reduced by approximately one order of magnitude with the proposed method compared to the classic front-tracking method, at a similar computational cost. 
In addition to the observed advantages of the NOA front-tracking method, its main limitation is inaccurate surface normal vector estimations in areas of very large interface curvature compared to the mesh resolution. An efficient and accurate solution approach for increasing the robustness of the method has been proposed in terms of a hybrid approach (HyNOA) in conjunction with the classic front-tracking method.

In the context of numerical interfacial flow modelling, the proposed method improves the overall accuracy and consistency of front-tracking methods by addressing two primary issues: 1) errors in volume conservation and shape preservation as results of frequent or continuous remeshing of the front, and 2) the advection of the front itself. The proposed NOA front-tracking method provides an accurate procedure to improve volume conservation and shape preservation during the advection of the front, at a computational expense similar to the classic front-tracking method.

\section*{Acknowledgements}
This research is funded by the Deutsche Forschungsgemeinschaft (DFG, German Research Foundation), grant number 420239128 and 458610925. Azur Hod\v{z}i\'c acknowledges the support from the European Research Council: This project has received funding from the European Research Council (ERC) under the European Unions Horizon 2020 research and innovation program (grant agreement No 803419). Fabien Evrard has received funding from the European Unions's Horizon 2020 research and innovation programme under the Marie Sk\l{}odowska-Curie grant agreement No 101026017. Clara M. Velte acknowledges the support from the Poul Due Jensen Foundation: Financial support from the Poul Due Jensen Foundation (Grundfos Foundation) for this research is gratefully acknowledged.

\end{document}